\documentclass[12pt,aps,preprint]{article}

\usepackage{setspace}
\usepackage{fullpage}
\usepackage{authblk}
\usepackage{hyperref}
\usepackage{appendix}
\usepackage{rotating, graphicx}
\usepackage{dcolumn}
\usepackage{multirow}
\usepackage{amsmath}
\usepackage{xcolor}
\usepackage{amsthm}
\usepackage{amssymb}
\usepackage{appendix}
\usepackage{subcaption}
\usepackage{eurosym}
\usepackage{dsfont}
\newcommand{\bi}{\begin{itemize}}
\newcommand{\ei}{\end{itemize}}

\newcommand{\be}{\begin{enumerate}}
\newcommand{\ee}{\end{enumerate}}

\newcommand{\beq}{\begin{equation}}
\newcommand{\eeq}{\end{equation}}

\newcommand{\beqnr}{\begin{eqnarray}}
\newcommand{\eeqnr}{\end{eqnarray}}

\theoremstyle{remark}

\title{The organization of the interbank network and how ECB unconventional measures affected the e-MID overnight market}

\begin{document}

\author[1,2,3]{Paolo Barucca} 
\author[1,4,5]{Fabrizio Lillo}
\affil[1]{\it Scuola Normale Superiore, piazza dei Cavalieri 7, 56126 Pisa, Italy}
\affil[2]{\it University of Zurich, Sch\"onberggasse 1
Zurich, 8001 Switzerland}
\affil[3]{\it London Institute for Mathematical Sciences, 35a South Street, London, UK}
\affil[4]{{\it QUANTLab, via Pietrasantina 123, 56122 Pisa, Italy}}
\affil[5]{{\it Department of Mathematics, University of Bologna, Piazza di Porta San Donato 5, 40126 Bologna, Italy}}
\date{\today}

\maketitle

\begin{abstract}
The topological properties of interbank networks have been discussed widely in the literature mainly because of their relevance for systemic risk. 
Here we propose to use the Stochastic Block Model to investigate and perform a model selection among several possible two block organizations of the network: these include bipartite, core-periphery, and modular structures. We apply our method to the e-MID interbank market in the period 2010-2014 and we show that in normal conditions the most likely network organization is a bipartite structure. In exceptional conditions, such as after LTRO, one of the most important unconventional measures by ECB at the beginning of 2012, the most likely structure becomes a random one and only in 2014 the e-MID market went back to a normal bipartite organization. By investigating the strategy of individual banks, we explore possible explanations and we show that the disappearance of many lending banks and the strategy switch of a very small set of banks from borrower to lender is likely at the origin of this structural change.
\end{abstract}

\section{Introduction}

The interbank market can be considered the plumbing of modern financial systems. Its functioning and proper organization is critical for the whole financial system and, as a consequence, for the economy as a whole. The recent global crisis has (once more) shown us that one of the first systems being affected by financial shocks is the interbank market who experienced a liquidity freeze or a substantial drop in turnover after the Lehman collapse \cite{georg}. The propagation of distress or default cascades in the interbank market has attracted a lot of attention and several models and metrics of bank's systemicness have been proposed.  For all these reasons, and thanks also to the increasing availability of suitable data, in the recent years many empirical studies have investigated the structure of the interbank network, which is the natural representation of the credit exposures among banks. In particular, analyses on different interbank networks have agreed on several "stylized fact" or statistical regularities frequently observed: very low connectivity, an heterogeneous degree distribution, low average distance between nodes, disassortative mixing, small clustering, and an heterogenous level of reciprocity \cite{Boss,iori2008network,frickelux2,iori2,amini2012,cont2011network}. 

Few studies have considered the large scale organization of the interbank network, i.e. how to divide it in homogenous functional groups. First of all, the interbank network is indeed a multilayer network, since different types of credit relations (characterized by the maturity or by the presence of collateral) are simultaneously present \cite{bargigli}. Second, when considering a single layer, i.e. one type of credit, the large scale organization that has been more frequently suggested is the core-periphery structure \cite{FrickeLux,craig,lelyveld,bargigli}. In plain words, the core is a subset of nodes which are maximally connected with other core members, while the periphery is the complementary subset made of nodes with no reciprocal connections, but connected to the network via core nodes. The inference of a core-periphery structure is a subtle issue, and several methods have been proposed (see Appendix \ref{why} for a discussion). Typically, a score function is chosen and then one finds numerically the partition of the nodes in two groups which maximizes the score. 

However all the proposed methods do not properly test for the alternative hypothesis that there is not such a structure, as well as they do not consider alternative hypotheses where the division in two groups exists, but it is not a core-periphery one. In this paper we propose to use the inference of the Stochastic Block Model (SBM) to identify the large scale structure of the interbank network. SBM has been introduced in \cite{borgatti,SBM1} and it is increasingly used in network modeling and in structure inference. The idea is that the set of nodes is partitioned in $m$ blocks (groups) and the probability that two nodes are linked depends on the groups they belong to. It is therefore the natural extension of the classical Erdos-Renyi \cite{erdos} random graph model when groups are present: within each block the subnetwork is of Erdos-Renyi type and between blocks the subnetwork is a random bipartite graph. 

When the SBM has two blocks several structures can be explained by the model, including the assortative mixing (modular structure), the disassortative mixing (bipartite structure), and the core-periphery structure. By performing a maximum likelihood  estimation of the SBM,  we can therefore compare the alternative structures and select the most likely one. Using a model selection technique we can also compare it with a purely random graph, i.e. a one block structure. The method also allows considering richer organizations, such as those composed by more blocks, even if in this paper we shall not consider them. Once the method for structural inference has been introduced, it is natural to investigate its dynamics, i.e. how it evolves in time and changes in periods of distress.

In this paper we apply our method to the Italian electronic market for interbank deposits (e-MID), previously investigated in many other papers (\cite{iori1,iori2008network,frickelux2,finger,FrickeLux,iori2,hatzo,kapar,temi} to name a few). Differently from them, however, we study the period 2010-2014, which for the Italian (and Euro) banking system has been a very critical period.  This is due to the sovereign debt crisis and to the response that the European Central Bank (ECB) gave to it. In particular at the beginning of 2012 the Long Term Refinancing Operations (LTROs) were launched with the specific aim of helping the distressed banking system. 

Our paper has three related research questions: (i) what is the large scale organization of the interbank network in "normal" periods? (ii) how did this organization change in response to the ECB unconventional measures? (iii) what change in the behavior of banks is responsible for this structural change? Our main findings can be summarized as follows. 
\begin{itemize}

\item First, in a normal period (e.g. in 2010 or in 2014) the most likely two block organization of the e-MID market is bipartite rather than core-periphery. This evidence is robust, since it is observed for weighted and unweighted networks and, even in the symmetrized version of the network, core-periphery is very infrequently observed (at least up to the monthly time scale). 

\item Second, in the two years after LTROs (2012-2013) the e-MID market experienced a dramatic change in structure. The typical bipartite structure connecting lenders and borrowers is not anymore the most likely inferred organization, but very often a random (structureless) organization better describes the data. As mentioned above, only in 2014 a partial recovery toward a typical two block structure is observed. 

\item Third, the micro origin of this organizational change can be understood by looking at how banks changed their strategy on e-MID around the LTROs. We observed that several banks, mostly characterized by a net lending profile, stopped trading in e-MID when LTROs were implemented. Moreover several of the remaining banks changed their strategy, with a prevalence of  large borrowing banks who became lending banks. In order to connect the macroscopic change in organization with the microscopic change in banks' behavior, we perform extensive numerical experiments showing that the changes described above of a very small number of banks can explain the emergence of the random structure observed in e-MID in 2012-2013. 

\end{itemize}





The prevalence of the bipartite over the core-periphery structure indicates that, at least for periods ranging from few days to a month, banks are in the (e-MID) market either to lend or to borrow. This is different from a view of the market where a core set of banks intermediates between banks in the periphery.

Some caveats on our findings are in order. Our empirical analysis is based on the e-Mid market, which is only a part of the full (Italian) interbank market, around the sovereign debt crisis. It is therefore possible that our results might be different in the whole market or in a different period. Concerning the first point, the fact that e-MID has been used in a very large of studies of interbank network, confirming results obtained in more complete datasets, encourage us on the significance of our results. It is true however that most of the above mentioned studies have investigated a period before the crisis, with a small overlap with ours. Therefore it might be possible that our result is specific of the investigated period. In any case we believe our most important contribution is to propose a rigorous statistical method  for inferring the best structure in the interbank network.

The paper is organized as follows. In Section \ref{sec:method} we introduce the inference technique which allows us to identify the large scale structure. In Section \ref{sec:data} we describe the investigated database, the e-MID market, and the timeline of critical events of the investigated period and in Section \ref{results} we present the results of this inference, showing, for different periods and different timescales, the most likely organization of the network, when a two block structure is inferred. We also document the change of network organization around LTROs. Section \ref{sec:unconventional} investigates in detail how individual banks reacted to LTRO, either by disappearing from the market or by changing the strategy from borrowing to lending (or, more rarely, viceversa). In Section \ref{simulations} we connect the change in organization with the change in banks' strategy by performing a sophisticated numerical exercise. Finally, in Section \ref{conclusions} we draw some conclusions and present some open issues.

\section{Statistical models for the block structure of a network}\label{sec:method}

We remind that a network can be represented by its weighted adjacency matrix $W=\{W_{ij}\}$. If the network describes the interbank market in a given time interval $[t,t+\Delta t]$, each node represents a bank and the element $W_{ij}$ is equal to the volume (in Euros) that bank $i$ lends to bank $j$ in this time interval. The matrix is clearly not symmetric. Moreover it is sometimes convenient to consider the binary adjacency matrix $A$ whose elements are $A_{ij}={\mathds 1}_{\{W_{ij}>0\}}$, where ${\mathds 1}_{B}$ is the indicator function of the set $B$. . 

There are different methods to identify the block structure (communities, core-periphery, etc) in a network. One way is to choose a score function depending on the assignment of nodes to blocks and then to find the assignment which minimizes or maximizes this score function. For example in community detection a widespread method is modularity maximization. A similar approach has been suggested for the identification of core-periphery  (see Appendix \ref{why} and discussion below around Eq. \ref{lipeq}). An alternative approach, followed here, is to specify a statistical model and then to infer the parameters of the model and the node assignments. One of the advantages of this latter approach is that it is not necessary to specify the sought structure, but possible alternatives can be simultaneously tested and selected.

In this paper we are interested in the large scale organization of the interbank networks, and for this reason we will consider as a statistical model the Stochastic Block Model (SBM) \cite{SBM1,SBM2} which includes different network organizations. Let us consider first a binary network, i.e. we discard the weights. The $N$ nodes are assigned to different communities by labelling each of them with a label $g_i$ ($i=1,...N$), representing the community to which node $i$ belongs. 
Then, for each couple of nodes $(i,j)$ belonging to the groups $(g_i,g_j)$, we add a directed link independently with probability $p_{g_ig_j}$ or not with probability $1-p_{g_ig_j}$ according to the entries of the so called \textit{affinity matrix} $p_{ab}$. Since the graph is directed, the affinity matrix $p$ is a $m\times m$ asymmetric matrix. Consequently, the model inference estimates $m^2+N$ independent parameters, i.e. the $m^2$ elements of the affinity matrix plus the $N$ assignments of each node to one of the $m$ groups. Given the assignment $g_i$ and $p$, we are defining an ensemble of graphs according to the probability mass function
\begin{equation}
\mathcal{P}(A | \boldsymbol{g},p)= \prod_{(i,j)}^N p_{g_ig_j}^{A_{ij}} (1-p_{g_ig_j})^{1-A_{ij}}.
\end{equation}

The affinity matrix characterizes the large scale structure of the network. Let us consider for simplicity the case $m=2$ in an undirected network. Depending on the relative size of the elements of the affinity matrix, one can describe the structure of the network:
\begin{itemize}
\item  If $p_{11} > p_{22} > p_{12}$ (or $p_{22} > p_{11} > p_{12}$) the network has a {\it modular structure}, i.e. two relatively isolated communities can be identified. Communities are sets of nodes much more strongly connected among themselves than with the rest of the network. The structure is more modular the smaller is $p_{12}$ with respect to $p_{11}$ and $p_{22}$.
\item If $p_{11} > p_{12} > p_{22}$ (or $p_{22} > p_{12} > p_{11}$) the network has a {\it core-periphery structure}. The nodes in group 1 (2) are strongly connected among themselves and those of group 2 (1) are poorly connected among themselves and connected to the network through links with nodes of group 1 (2).
\item If $p_{12} > p_{11} > p_{22}$ (or $p_{12} > p_{22} > p_{11}$) the network has a {\it bipartite structure}. Links are preferentially observed between two nodes belonging to two different groups, while links between nodes belonging to the same group are rarer.
\end{itemize}

Note that a SBM with all $p$s different from zero is not {\it exactly} modular or core-periphery or bipartite. In fact, a perfect modular structure is one where $p_{12}=0$ or a perfect bipartite structure is one where $p_{11}=p_{22}=0$. Thus the ranking of the $p$s gives the best representation of the two block network organization. 

It is important to point out that SBM identifies groups according to the probabilities of a link within a group or between groups, rather than considering the number of links. A clear manifestation of this difference can be highlighted by considering a core-periphery SBM. A standard method to identify the core (see \cite{FrickeLux,lip}) is to  find the set $S_1$ of nodes which minimizes the the score function
\begin{equation}\label{lipeq}
Z(S_1)=\sum_{(i<j)\in S_1} {\mathds 1}_{\{A_{ij}=0\}}+\sum_{(i<j)\not\in S_1} {\mathds 1}_{\{A_{ij}=1\}}.
\end{equation}
Absence of links in the (putative) core is penalized as well as presence of links in the (putative) periphery. It is possible to show that the expected value of this score function over an ensemble of core periphery SBMs with fixed parameters and nodes assignments does not return the correct (according to SBM) core size, but underestimates it significantly and this effect is stronger the larger is the network. We believe that a method which considers probabilities rather than numbers of links is more appropriate for modeling block organization and for this reason we propose the SBM as a valid tool.

The SBM model can be generalized to the case where links can assume arbitrary weights $W_{ij}\in \mathbb{N}$. The probability mass function becomes
\begin{equation}\label{eq:ML}
\mathcal{P}(W | \boldsymbol{g},p)= \prod_{(i,j)}^N \frac{p_{g_ig_j}^{W_{ij}}}{W_{ij}!}\exp{(-p_{g_ig_j})}.
\end{equation}
Since $W_{ij}$ is an integer, it can be interpreted as the presence of $W_{ij}$ links between node $i$ and $j$. This is the standard way of dealing with an ensemble of weighted graphs, by interpreting a weighted link as a multiple link. In our analysis of the interbank market we discretize the volume $V_{ij}$ exchanged in an interbank transaction by using  $W_{ij}=\left \lfloor{\log_c(1+V_{ij})}\right \rfloor$, where $c=1+\min\{V_{ij} | V_{ij}>0\}$ and $\lfloor..\rfloor$ is the floor function. 
Thus for the unweighted case $p_{g_ig_j}$ is the probability of a link between a node in group $g_i$ and a node in group $g_j$, while in the weighted case it is the average weight between the two nodes. 

Less flexible generative models can be constructed by constraining each node to have a fixed degree with the nodes of each different group (Regular Stochastic Block Model (r-SBM))\cite{SBM2} or by constraining each group to have a fixed number of links within itself and with all different communities \cite{Tiago1}, the so-called micro-canonical Stochastic Block Model, that is the one we used for the analysis of the e-MID interbank network (see below).

\subsection{Inference of the SBM}

Inferring a SBM means to find the parameters $(\boldsymbol{g}, p)$ given an observed network. Despite the fact that writing the likelihood of an SBM is straightforward, its maximization might be hard and therefore several different inference methods have been proposed \cite{SBM2,Tiago1,Tiago2,Tiago3}.
Here we consider a recently proposed method \cite{Tiago3} which is based on the application of Markov Chain Monte Carlo (MCMC) to a slightly different version of the above described model.

Consider $N$ nodes divided in $m$ groups of size $\{n_r\}_{r=1}^m$ and a fixed number of intergroup links $\{e_{rs}\}_{r,s=1}^m$, between each couple of groups $(r,s)$.
It is thus possible to compute, in the large $N$ limit, the total number $\Omega$ of possible graphs compatible with the constraints $\{n_r\}$ and $\{e_{rs}\}$ as \cite{Tiago1}: 
\begin{equation}
\Omega = \exp(\sum_{r,s}n_rn_s\mathcal{H}(\frac{e_{rs}}{n_rn_s}))
\end{equation}
where $\mathcal{H}(x) = -x\ln(x)-(1-x)\ln(1-x)$.
If we now consider the microcanonical entropy $S = \ln(\Omega)$ we then look for the assignment $\bold{g}$ in $m$ blocks that minimizes it. It can be shown that this is  equivalent to maximizing the log-likelihood of the assignment in the microcanonical SBM ensemble. 
The algorithm to minimize $S$ has been thoroughly studied in \cite{Tiago3} and it is a MCMC integrated with two specific heuristics to optimize the choice of nodes moves from one block to another and to avoid local minima in the entropy. 
 
When the number of blocks is unknown it is necessary to estimate the cost of introducing extra-parameters for the description of the graph in order not to converge to a trivial assignment where each node is assigned to a different block. 
The choice of the number of blocks is addressed in \cite{Tiago2} through the introduction of $\mathcal{L}$ that measures the volume of the space of parameters in the SBM with a given number of blocks $m$ and edges $K$. 
The sum of the entropy $S$ and $\mathcal{L}$ defines the description length $\Sigma(\bold{g},m,K)$ \cite{Tiago2}. The complete inference algorithm based on the minimization of $\Sigma$ that has been used in the present work can be found on \href{https://graph-tool.skewed.de}{\textit{graph-tool.skewed.de}} together with a detailed documentation.

\section{Data}\label{sec:data}

\begin{figure}[t]
\centering
\includegraphics[width=90mm,angle=0]{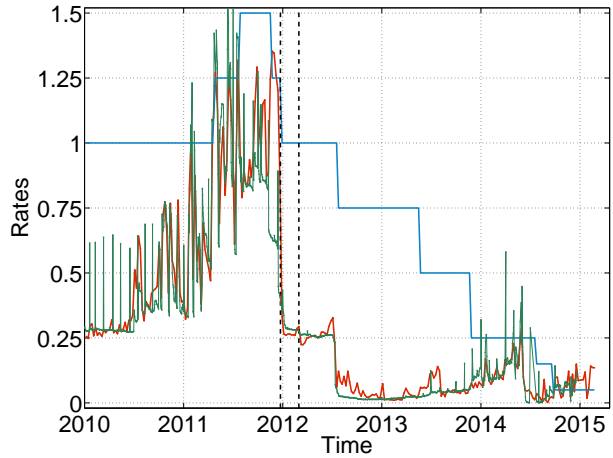}		
\caption{Average weekly rate of overnight lending at e-MID in the period 2010-2014 (red line) compared with the ECB fixed rate (blue line), and overnight LIBOR based on Euro (green line). The vertical dashed lines indicate the two LTRO measures by ECB on Dec. 22, 2011 and Feb. 29, 2012.}
\label{rates}
\end{figure}

\begin{figure}[t]
\centering
\includegraphics[width=80mm,angle=0]{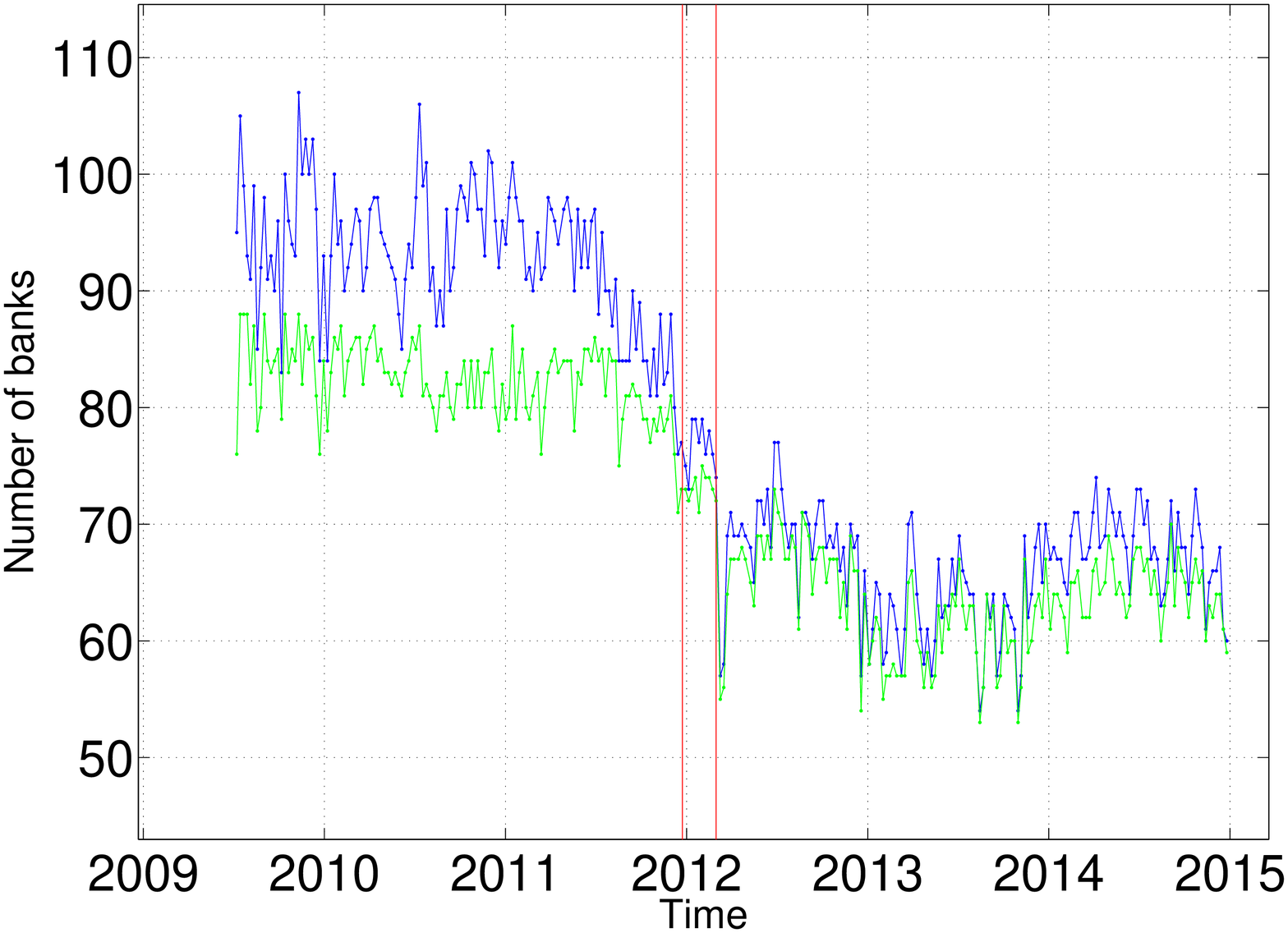}
\includegraphics[width=80mm,angle=0]{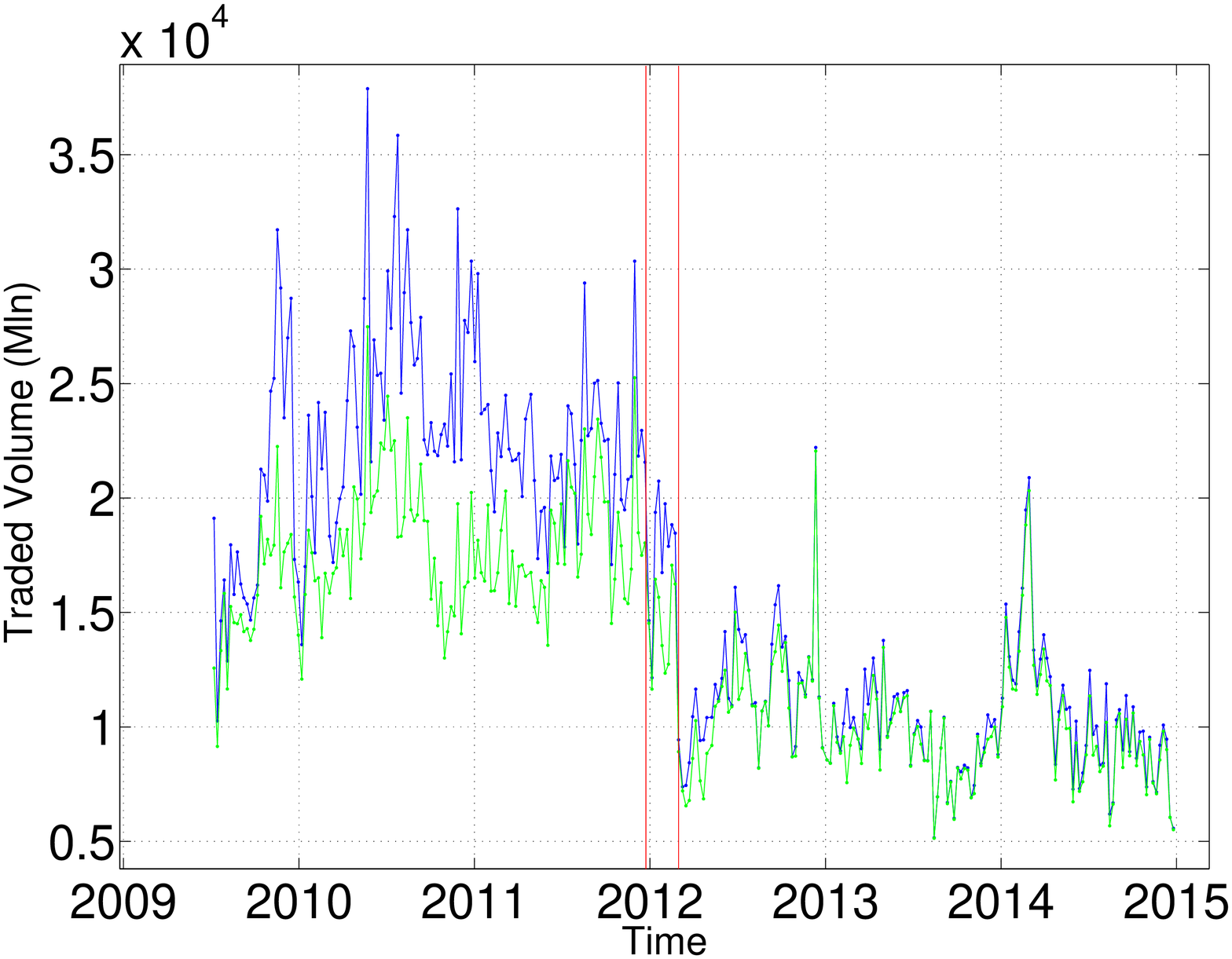}		
\caption{Weekly number of banks (left) and weekly traded volume in Million of Euros (right) traded at e-MID from June 2009 to December 2014. The blue line refers to all the banks while the green line refers only to Italian banks. Vertical red lines indicate the two LTRO measures by ECB on Dec. 22, 2011 and Feb. 29, 2012.}
\label{bankts}
\end{figure}

The Italian electronic market for interbank deposits (e-MID) is a screen-based platform for trading of unsecured money-market deposits operating in Milan. Average daily trading volumes were $3.6$ Billions of Euro in 2010 to $2.0$ in 2014. 
The dataset includes e-MID transactions from July 2009 to December 2014. We focus on the Italian banks constituting roughly 68\% of the banks and 84\% of the total volume in the whole period (see however discussion on Fig. \ref{bankts} for the time series of these quantities). Moreover we focus on overnight exposures. For each transaction we have the information on the coded identity of the borrower and of the lender, the begin and end date of the credit, and its amount (in Euro).

The five year investigated period (2010-2014) was quite complex for the Italian and European banking system, since it overlaps significantly with the Euro-zone sovereign debt crisis. The second part of 2011 witnessed a rapid worsening of the financial crisis and a surge in global uncertainty. In response to this critical period, on December 8, 2011 ECB announced the two 3-year Long Term Refinancing Operations (LTROs) that took place on the 22nd of December 2011 and on the 29th of February 2012 for a total amount of about 1.03 trillions. LTROs are a cheap loan schemes for European banks  and were implemented with the aim of increasing cash flow and avoiding a severe credit crunch. Thus LTROs helped banks to face liquidity problems due, among other things, to their difficulty of getting back the money lent to investors. Banking signing up for LTRO asked the ECB for the loan, backing it with a collateral. In December Italian banks took Û110 billion Euro. 

Even if the focus of the paper is on the behavior of e-MID around LTROs, it is useful to remember that the investigated period overlaps with other ECB unconventional measures. These includes the Securities Market Program (SMP) on May 10, 2010 and on August  2, 2011 and the Outright Monetary Transactions (OMT) on August 2, 2012. Finally, it is worth mentioning the famous speech of Mario Draghi on July 26, 2012 at the Global Investment Conference in London\footnote{"The ECB is ready to do whatever it takes to preserve the Euro. And believe me, it will be enough."}

Figure \ref{rates} shows the average weekly rate of overnight lending at e-MID in the period 2010-2014 and compares it with the ECB fixed rate and with the overnight LIBOR based on Euro. The pre-LTRO period (2010-2011) was characterized by a rapid surge of the rate, which experienced a fivefold increase in less than two years. The first LTRO caused a drop of the rate to the values of the beginning of 2010 and the decrease of the e-MID rates is much higher than the concurrent decrease of the ECB rate, indicating an indirect effect of LTRO on e-MID rates. Interestingly the second LTRO had a much smaller effect on rates. We observe another very significant drop in rates around the second week of July of 2012 when ECB lowered significantly the interest rates. Similarly other downward jumps of the rate later on are in correspondence with ECB interventions on the interest rate.

Thus the e-MID market was strongly affected by the European sovereign debt crisis as well as by the unconventional monetary measures adopted by the ECB. The latter, in fact, strongly affected the funding capability of Italian banks and therefore their activity in the interbank market, as the e-MID. A first exploratory analysis confirms this statement. Left panel of Figure \ref{bankts} plots the number of active banks in each week of the investigated period. It is evident that the number of banks dropped from more than 90 in 2010 and 2011 to roughly 60 in 2013 (and slightly more in 2014). Also the traded volume (right panel of Figure \ref{bankts}) shows a similar drop. It is important to notice that this change occurred simultaneously to the LTROs (red vertical lines). Note also that the number of foreign banks starts declining {\it before} the first LTRO, i.e. during the second half of 2011, and when the the first LTRO started the number of foreign banks (the difference between the blue and the green line in the left panel of Figure \ref{bankts}) became very small. 

\begin{figure}[t]
\centering
\includegraphics[width=80mm,angle=0]{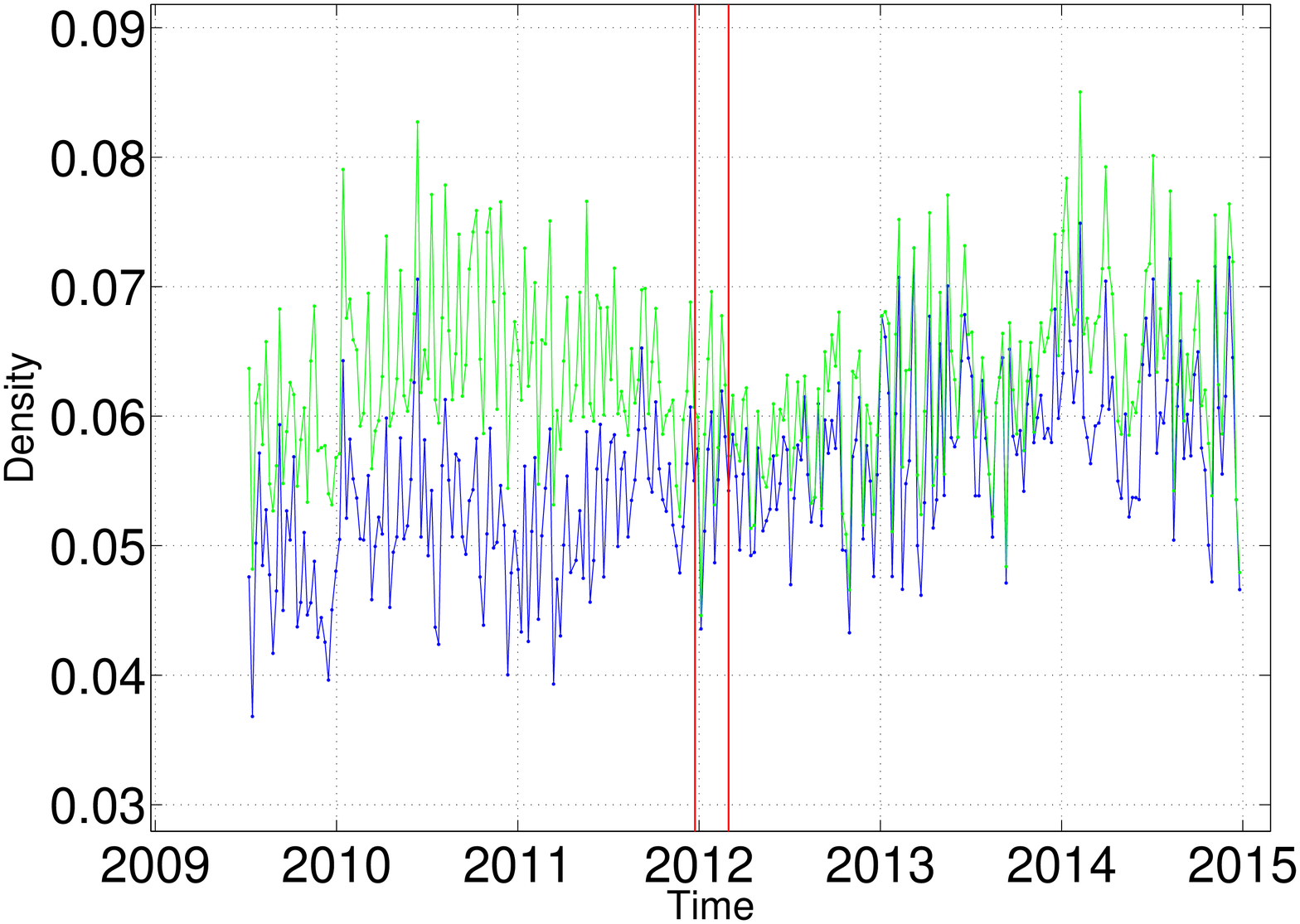}
\includegraphics[width=80mm,angle=0]{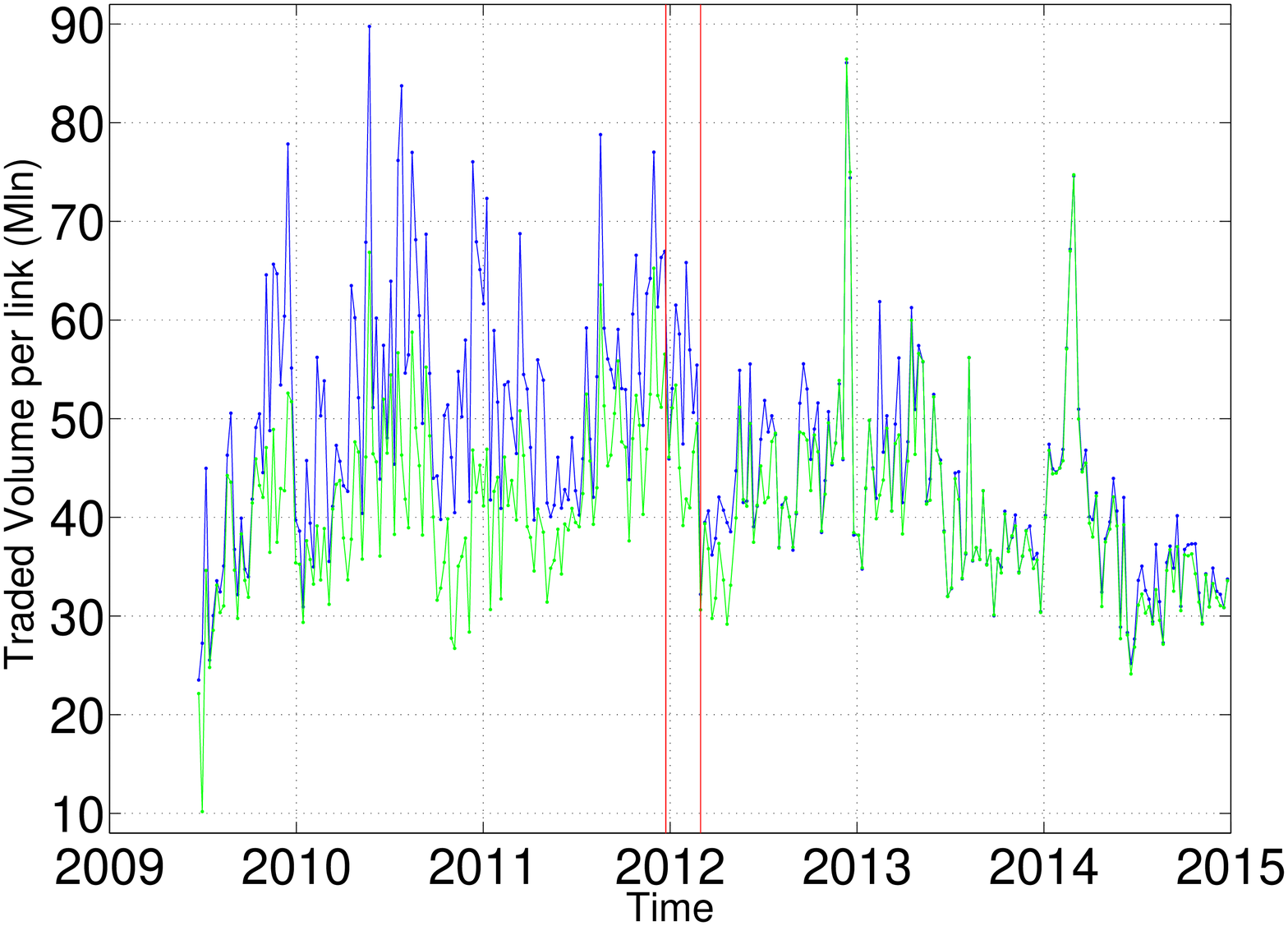}
\caption{Left. Weekly density (i.e. number of links divided by the total number of possible links) of e-MID from June 2009 to December 2014. Right. Weekly volume per link (i.e. total volume in Million of Euros divided by the number of links. The blue line refers to all the banks while the green line refers only to Italian banks. Vertical red lines indicate the two LTRO measures by ECB on Dec. 22, 2011 and Feb. 29, 2012.}
\label{bankdensts}
\end{figure}

The change in the number of banks and traded volumes does not imply that the structure and characteristics of the interbank network changed in response to the unconventional measures. For example, the density of the network (i.e. the number of links over the total number of possible links) did not display a similar transition (see the left panel of Fig. \ref{bankdensts}). This means that the banks that remained in the market after the LTRO did not replace the missing opportunity of trading with banks who had disappeared with new links with banks they were not connected to. Rather, either they looked for other funding/lending opportunity outside e-MID or they increased the amount exchanged with the banks they were trading before LTRO and that did not leave e-MID. In order to discriminate between these two alternatives, the right panel of Fig. \ref{bankdensts} shows the time series of the average amount exchanged per link. The series does not show a significant increase around the time of the LTROs, indicating that remaining banks did not increase the volume of trading with the remaining banks, but either trade less in the interbank market or trade outside e-MID. With the available data we are not able distinguish these two cases. 

Despite the fact that density has not changed, other organizational characteristics of the network might have changed. In order to investigate this important point, in the next Section we apply the inference of the SBM to the e-MID interbank network and we perform model selection among different two-block structures.

\section{The large scale organization of e-MID market}\label{results}

In this Section we report the results of the inference of the large scale structure of the e-MID interbank network. A key variable for the analysis is the time scale of aggregation of the transactions. Even if we have data with daily resolution, it is interesting to investigate how the structure of the interbank network changes at different aggregation scales, ranging from daily to monthly time scale \cite{finger}. In fact, while at the daily time scale we might observe fluctuations related to the contingent excess or lack of liquidity, by considering longer time scales we might be able to identify long run tendencies in the structure of the bank's credit relations. In the following we aggregate all the transactions between day $t$ and $t+\Delta t$, where $\Delta t=1,5,20$ days, corresponding to daily, weekly, and monthly  time scales. In all cases we will use non overlapping time windows.

As discussed, we infer the SBM with $m=2$ in order to identify which two block structure among modular, core-periphery, and bipartite best describes the e-MID at different time scales. It is important to remember that the algorithm performs a model selection considering also alternative models with less than $m$ blocks. Thus in the $m=2$ inference we will consider also the case where no block structure is identifiable, i.e. the network is better described by a random graph (Erdos-Renyi). 


According to what described above, the inference with $m=2$ blocks can give four different results on the best structure describing the data. First, the method can select between a one block (random graph) or two block structure\footnote{Technically, when the inference gives a large block and a very small one (i.e. composed by less than 5\% of the nodes, typically 2 or 3 banks), we shall consider it equivalent to a one block structure.}. In this second case, depending on the relative size of the parameters $p_{ij}$ of the affinity matrix, the method selects among the modular, core-periphery, or bipartite structure. 
Table \ref{tab2block} reports the fraction of times the four different structures are identified at different time scales. We have considered both the case of the inference on the unweighted (top) and on the weighted (bottom) network. 

\begin{table}
\begin{center}
\begin{tabular}{l | rrrr | rrrr| rrrr|}
  \hline
\multirow{1}{*}{Unweighted} &  \multicolumn{4}{c}{Day} &  \multicolumn{4}{c}{Week} &  \multicolumn{4}{c}{Month}  \\
\hline
Year & B & C & M &R & B & C & M &R & B & C & M &R \\
\hline
2010 & 53&0&0&47&92&0&0&8&100&0&0&0\\
2011 & 55&0&0&45&90&0&0&10&100&0&0&0\\
2012 & 41&0&0&59&55&0&0&45&75&0&0&25\\
2013 & 38&0&0&62&34&0&0&66&58&0&0&42\\
2014 & 53&0&0&47&80&0&0&20&100&0&0&0\\
 \hline
 \end{tabular}\\ 
  \begin{tabular}{l | rrrr | rrrr| rrrr|}
  \hline
   \hline
\multirow{1}{*}{Weighted~~~} &  \multicolumn{4}{c}{Day} &  \multicolumn{4}{c}{Week} &  \multicolumn{4}{c}{Month}  \\
\hline
Year & B & C & M &R & B & C & M &R & B & C & M &R \\
\hline
2010 & 53&0&6&41&92&0&0&8&100&0&0&0\\
2011 & 55&0&5&40&90&0&0&10&100&0&0&0\\
2012 & 41&0&9&50&55&0&0&45&75&0&0&25\\
2013 & 39&0&5&56&36&0&2&62&58&0&0&42\\
2014 & 52&0&4&44&80&0&0&20&100&0&0&0\\
\hline
 \end{tabular}
   \caption{Percentages of inferred structures in the e-MID interbank market at different temporal aggregations in the 5 investigated years. The structures are bipartite (B), core-periphery (C), modular (M), and no structure (R). The table refers to the directed unweighted (top) and weighted (bottom) case.}
\label{tab2block}
\end{center}
\end{table}

The first observation is that the method almost always infers either a bipartite structure or a one block structure. The  core-periphery structure is never found and the modular structure is found only few percent of the times in the daily weighted case. While the latter observation is somewhat expected, the former one comes to a surprise, since many empirical studies have documented a core-periphery organization of the interbank network \cite{FrickeLux,craig,lelyveld}, including in the e-MID. We argue that this is due to the methods other papers use to infer the structure and in particular to the lack of comparison with other alternative organizations. Other elements, for example discarding the information on link directionality or weight, also make more likely the identification of a core-periphery structure. In the next section we present an extensive analysis on this point and we bring more evidences supporting our claims.

\begin{table}
\begin{center}
  \begin{tabular}[width=4cm]{  l | r | r | r p{1cm} |}
        & B & L  \\ \hline
    B& $1.16~(0.61,1.52)$ & $0.20~(0.0, 2.2)$ \\ 
    L& $23~(21,24) $ & $3.70~(2.07,4.8)$ \\ 
   \end{tabular}
\hspace{2cm}
  \begin{tabular}[width=4cm]{  l | r | r | r p{1cm} |}
     & B & L  \\ \hline
     B& $1.02 \pm 0.96 $ & $0.071 \pm 0.107$ \\ 
    L& $8.26 \pm 2.09 $ & $0.33 \pm 0.24$ \\ 
   \end{tabular}
   \caption{Left. Average affinity matrix (in percentage) of the weekly interbank network when a bipartite structure is found. Numbers in parenthesis are the first and third quartile. Right. Average total weight (in Billions of Euros) between groups  when a bipartite structure is found in weekly networks of 2014.}
\label{affinity2}
\end{center}
\end{table}

Interestingly the inference on the weighted or on the unweighted case gives very similar results so in the following we focus on the latter case. Focusing for simplicity on the weekly time scale, we observe that the typical size of the two groups, when the bipartite structure is found, is quite stable in time: the group of borrowers is composed by $45\pm 9$ banks while the group of lenders is composed by $30\pm 8$ banks. 
The average affinity matrix, when a bipartite structure is found, is shown in Table \ref{affinity2}. The values of the affinity matrix give the probability of a link between two nodes in the two identified groups. For example, the value $1.16$ reported in the BB element of the affinity matrix means that on average (across the weeks where a two block structure is inferred) the probability of a link between two borrower banks is $1.16\%$. Since we have a value of the affinity matrix for each week, we can perform a t-test on the significance of the ranking between the elements of the matrix. We find that with a $p<10^{-16}$ the following relations hold

$$p_{LB}>p_{LL}>p_{BB}>p_{BL}$$

Since the difference in the values of the elements of the affinity matrix is statistically significant, we conclude that at the weekly time scale the structure of the interbank network is clearly bipartite and not core-periphery or modular\footnote{We have also performed extensive numerical simulations on the role of small sample size (e.g. number of banks) in the inference. By simulating a SBM with the inferred parameters and the same number of banks, we found that the inference method is able to identify very clearly the model even when the sample size is small (data available upon request).}. The right part of Table \ref{affinity2} shows the average total weight between groups, i.e. the average weekly  credit exchanged, in 2014. Also by looking at this quantity it is clear that the largest amount of credit is observed from group L to group B.


Considering the dynamics of the network organization, Table  \ref{tab2block} shows a strikingly different behavior between the years before  LTRO (2010-2011)  and those immediately after it (2012-2013). In the first period the bipartite structure is much more frequently observed, while after LTRO a random structure becomes significantly more frequent. This effect is stronger at weekly and monthly time scale, while the more volatile structure of daily networks makes the random structure globally more frequent. Interestingly in 2014 the interbank network has a structure closer to the pre-LTRO period.   

It is therefore intriguing to ask whether LTRO was responsible for this change of organization and in the following we will try to answer this question. We can rule out the possibility that this change of organization is due to the disappearance of banks and links. To show this we ran numerical simulations where we generate artificial networks according to the SBM with parameters and number of banks equal to those observed in 2010, when a clear bipartite structure was identified. We then randomly remove banks and the corresponding links to numbers similar to those observed in 2012 and we perform the inference of the two-block structure. In the $1,000$ simulations we never observe a case when the inference was not able to identify the bipartite as the most likely structure. This means that the drop in the number of banks alone cannot be at the origin of the change of interbank network organization due to LTRO. In other words, our observation is not a statistical artefact due to small samples. In Section \ref{simulations} below we explore with more sophisticated numerical experiments the possible causes of this change of organization.

In conclusion this extraordinary policy measure not only lead to a shrinkage of the volume and number of banks in the e-MID market, but it also changed dramatically its large scale organization. In section \ref{sec:unconventional} we investigate the possible causes of the structural change

\subsection{Core-periphery vs bipartite structure}\label{comp}

We have seen that at all scales of temporal aggregation the e-MID interbank network displays a strong directed structure, meaning both that reciprocity of links is very low and that the difference between nodes out- and in-degree is far from zero for most nodes. These two aspects, already outlined in previous works, reveal that the e-MID market is a direct and polarized market where banks fulfil specific needs for borrowing or lending money. 

Previous literature \cite{FrickeLux,craig} on the structure of the interbank network have focused their attention on the identification of a core-periphery structure. Here we discuss possible origins of this discrepancy and how different methods and/or data transformation can affect the inferred structure.

There are three main reasons why different structures can be inferred in the interbank network, namely (i) symmetrization and binarization of the network, (ii) time aggregation, and (iii) choice of the inference method,

\begin{table}
\begin{center}
\begin{tabular}{l | rrrr | rrrr| rrrr|}
  \hline
\multirow{1}{*}{Unweighted~~~Symmetric} &  \multicolumn{4}{c}{Day} &  \multicolumn{4}{c}{Week} &  \multicolumn{4}{c}{Month}  \\
\hline
Year & B & C & M &R & B & C & M &R & B & C & M &R \\
\hline
2010 & 53&1&8&39&86&6&0&8&83&17&0&0\\
2011 & 54&1&7&39&73&18&0&10&58&42&0&0\\
2012 & 40&0&6&53&47&8&0&45&50&25&0&25\\
2013 & 37&1&4&58&32&2&2&64&8&50&0&42\\
2014 & 52&0&4&44&74&6&0&20&42&58&0&0\\
 \hline
 \end{tabular}\\ 
\begin{tabular}{l | rrrr | rrrr| rrrr|}
  \hline
\multirow{1}{*}{Weighted~~~Symmetric} &  \multicolumn{4}{c}{Day} &  \multicolumn{4}{c}{Week} &  \multicolumn{4}{c}{Month}  \\
\hline
Year & B & C & M &R & B & C & M &R & B & C & M &R \\
\hline
2010 & 37&1&47&15&84&6&4&6&58&42&0&0\\
2011 & 38&7&39&16&57&31&2&10&67&33&0&0\\
2012 & 40&1&48&11&45&10&22&24&50&25&0&25\\
2013 & 36&1&50&13&24&10&44&22&25&33&25&17\\
2014 & 42&2&41&15&72&8&6&14&75&25&0&0\\
\hline
 \end{tabular}
   \caption{Percentages of inferred structures in the e-MID interbank market at different levels of aggregation in the 5 investigated years. The structures are bipartite (B), core-periphery (C), modular (M), and no structure (R). The table refers to the symmetrized unweighted (top) and weighted (bottom) case.}
\label{tab2blocksymm}
\end{center}
\end{table}

First of all, core-periphery might be more easily identified under symmetrization, since passing from a directed to an undirected network, information about the structure is lost and structure identification may be strongly influenced by total degree. Binarization, i.e. neglecting the weights of the links and considering only the topological structure, enhances even more the probability of identifying a core-periphery structure. Table \ref{tab2blocksymm} shows the results of the SBM inference under symmetrization and neglecting (top) or considering (bottom) weights. The comparison with Table \ref{tab2block} partly confirms the trends toward a larger probability of inferring a core-periphery structure\footnote{It is also worth mentioning that most algorithms on networks, including inference of SBM and other centrality metrics, as for instance eigenvector centrality, are highly dependent on degree-heterogeneity. If not properly taken into account \cite{SBM1,Tiago1,barucca}, this heterogeneity affects structure inference. In fact a core-periphery structure may be found where the highest degree nodes are put in the core and lowest degree nodes are put in the periphery without the core being more densely connected than the periphery.}. 

Secondly core-periphery can arise from time-aggregation, nodes may change their behavior in time and start having both out- and in-connections thus making the structure less bipartite and eventually the network becomes core-periphery. Previous works in interbank markets \cite{FrickeLux,craig,iori2008network} have focused on monthly and quarterly aggregated networks that are denser than daily and weekly networks and present a less clear bipartite structure: also banks with a clear block-membership, as borrower or lender, may eventually perform a transaction of opposite sign and thus become harder to assign to a bipartite structure. We empirically observe how symmetrization enhances core-periphery structure on long time-scales while on shorter time-scales the network remains highly bipartite in the generalized inference framework of SBM (see Table \ref{tab2blocksymm}). On quarterly time scales we infer no core-periphery structure on the $20$ quarterly aggregated asymmetric matrices while in the symmetrized case we infer a core-periphery structure in $16$ quarters out of $20$.

Thirdly there is a methodological reason. Many algorithms \cite{boyd,lip} {\it assume} the existence of a core-periphery organization a-priori and find the best core-periphery without testing for its presence against alternative structures. Even though it might be useful to quantify centrality and coreness measures in networks per-se, when it comes to the interpretation of results it is important to assess the relative relevance of a given structure with an appropriate statistical model, as in the case of SBM inference where the significance of different structures is quantitatively compared through the log-likelihood.  It is worth mentioning that, in a companion paper \cite{barucca} we apply a different inference method (belief propagation) for SBM to the symmetrized interbank network. We find results substantially consistent with those presented here, confirming the robustness of our findings to the inference method.

\section{How banks reacted to ECB unconventional measures} \label{sec:unconventional}

In order to understand the change in organization of the interbank market around LTROs, we investigate possible changes of strategy of individual banks. For simplicity the strategy of a bank is here considered as the prevalence of borrowing or lending activity in e-MID in a given time period. To monitor the strategy and its change we consider the weighted matrix $W(t)$ at day $t$, whose elements $W_{ij}(t)$ equals the amount of the loan from bank $i$ to bank $j$. We then introduce the net e-MID balance of bank $i$ in a given day $t$ as  $\Delta b_i(t) = \sum_{j} W_{ij}(t) - W_{ji}(t)$. For each bank we evaluate $b_i(t) = \sum_{s=1}^t \Delta b_i(s)$, that is the net amount of lent (when the sign is positive) or borrowed (when the sign is negative) money in the e-MID interbank market. 

In order to compare banks of different size of trading activity we divide the quantity $b_i(t)$ by the maximum absolute value achieved by it during the investigated period, i.e. the normalized inventory is $\tilde b_i(t)=b_i(t)/B_i$, where $B_i=\max_t |b_i(t)|$. In this way for each bank $\tilde b_i(t)$ is bounded in $[-1,1]$, typically reaching only one of these boundaries. A locally positive (negative) slope of $\tilde b_i(t)$ indicates a lending (borrowing) activity, while when $\tilde b_i(t)$ is constant the bank is very likely inactive in e-MID. 

\begin{figure}
\centering
\includegraphics[width=7cm]{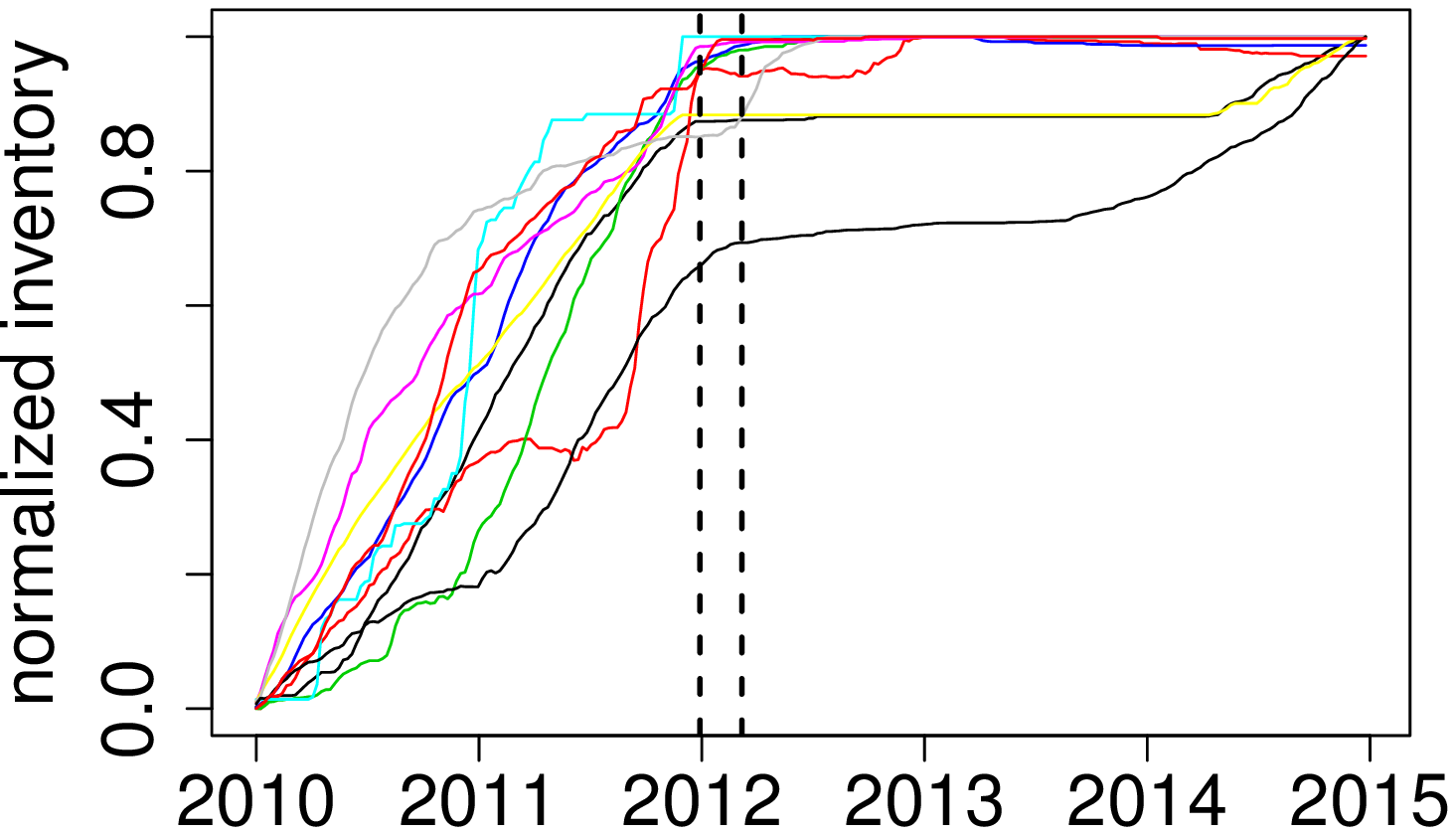}
\includegraphics[width=7cm]{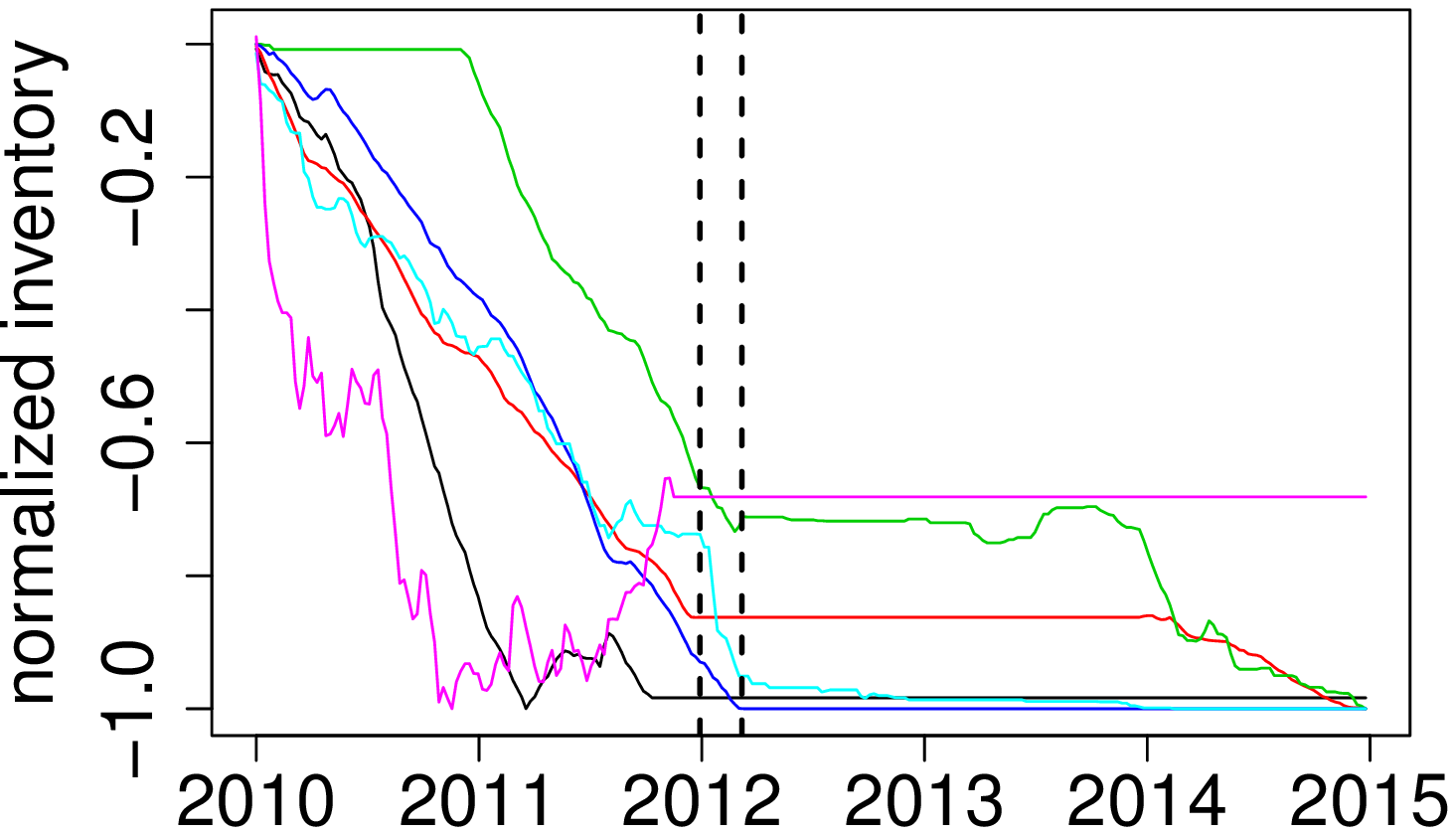}\\
\includegraphics[width=7cm]{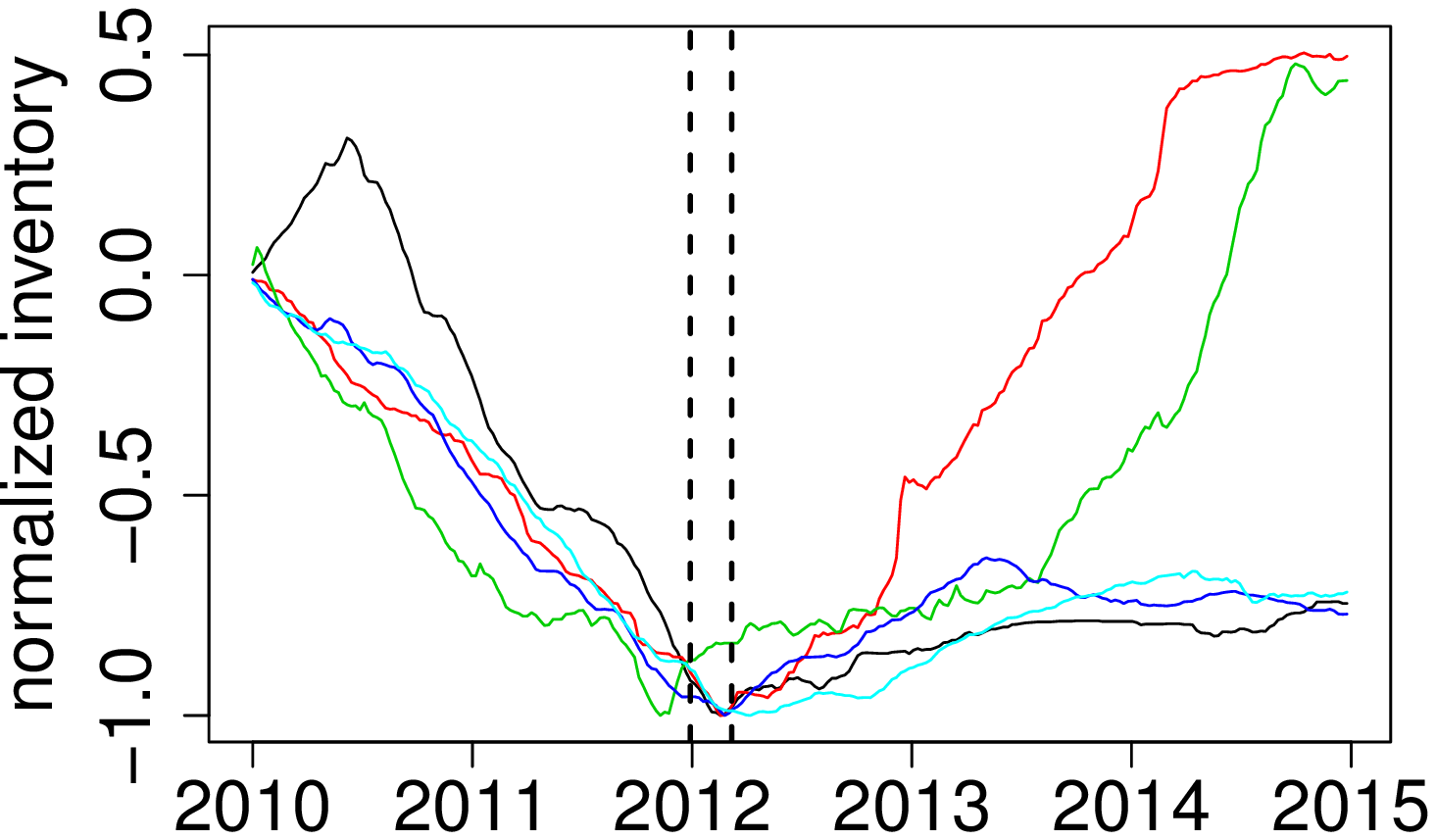}
\caption{Normalized inventory $\tilde b_i(t)$ of three groups of banks in the e-MID market. The top left (right) shows $10$ ($6$) lending (borrowing) banks which stopped trading at the time of LTROs (indicated by two vertical dashed lines). The bottom panel shows $5$ banks that switched from a borrowing to a lending behavior. Note that the scale of the y axis is different in the three panels to improve their readability.}
\label{inventory}
\end{figure}

Figure \ref{inventory} shows the normalized inventory $\tilde{b_i}(t)$ for many selected banks. Specifically, in the top left panel we show $\tilde{b_i}(t)$ for $10$ banks which in the pre-LTRO period were typically lender and stopped the trading activity in e-MID at the time of LTRO. Similarly in the right panel we show a sample of $6$ borrowing banks. Finally in the bottom panel of Figure \ref{inventory}  we show the normalized inventory of $5$ banks which were borrower before the LTRO and abruptly switched to a lender strategy at the time of the LTRO. We observe one or two banks that switched the strategy in the opposite direction, i.e. from lending to borrowing (data not shown). In all the examined cases it is astonishing how sharp is the regime switch in the banks' strategy at the time of the LTROs.


We then performed a more systematic analysis of the changes of strategies of the banks. In order to take into account both the size and the borrowing/lending profile of banks in the investigated period, at each time $t$ we construct five non-overlapping sets: (i) Big borrowers ($BB_t$): banks with a net-balance greater (in absolute value) than the median of the net-balance distribution among the borrowers. (ii) Small borrowers ($SB_t$): banks with a net-balance smaller (in absolute value) than the median of the net-balance distribution among the borrowers. (iii) Big lenders ($BL_t$): banks with a net-balance greater (in absolute value) than the median of the net-balance distribution among the lenders. Small lenders  ($SL_t$): banks with a net-balance smaller (in absolute value) than the median of the net-balance distribution among the lenders. Finally we define the Non Active ($NA_t$) set, composed by banks who did not trade at time $t$. 

In order to investigate the change of strategy around LTROs, we estimated the $5\times 5$ transition matrix between the first quarter of 2011, which is relatively far from the most critical period of the Italian sovereign debt crisis, and the second quarter of 2012, which falls immediately after the second LTRO. The matrix is shown in Table \ref{transition}.
Apart from the diagonal terms, indicating banks that preserved their strategy, we observe that $15/51\simeq 29\%$ of the lender banks, but only $5/21\simeq 19\%$ of the borrower banks become inactive. This is consistent with the behavior observed in the top left panel of Fig. \ref{inventory}. Moreover $6/16\simeq 38\%$ of the BBs in 2011 became lenders after LTRO, while only $4/26\simeq 15\%$ of the BLs became borrowers. This asymmetry is evident also by considering all borrowers and lenders. Finally among the banks that were inactive in 2011, $18$ became lenders, while only $9$ became borrowers. Thus the LTRO favoured the arrival of new lenders. 
 
The above analysis shows that a considerable fraction of banks changed completely their strategy in the e-MID interbank market as a consequence of the LTRO measures. In particular some banks (more lending than borrowing banks) stopped their activity on e-MID. The place of the missing lending banks was mostly taken by the banks who switched from borrowing to lending, even if a small number of new lenders stepped into the market. It is reasonable to expect that the new and cheap funding provided by the LTRO allowed typical borrowers to fund themselves outside the market. Moreover the switching behavior suggests that some of them used this opportunity also to lend money in the e-MID to other banks, which probably did not have access to the LTRO program. This activity was probably pursued also by banks which were not active in the e-MID before 2012, but used the LTRO as an opportunity to become lenders. This flow of liquidity, triggered by the exceptional measures, wiped out from e-MID a large fraction of traditional lenders.  

In conclusion there is a strong indication that the change in the large scale organization in 2012-2013 that we have documented above is related to this change in banks behavior. Although proving causality relations is typically very hard, in the next Section we create an explicit link between the banks' behavioral change at the micro level and the large scale organization of the network at the macro level.

\begin{table}
\begin{center}
\begin{tabular}{|l | rrrrr |r|}
\hline
 & $BB_t$ & $SB_t$ & $SL_t$ & $BL_t$ & $NA_t$ &\# of banks\\
\hline
$BB_{t-1}$ & 37&21&26&5&11&19\\
$SB_{t-1}$ & 5&55&5&20&15&20\\
$SL_{t-1}$ & 23&0&35&15&27&26\\
$BL_{t-1}$ & 12&4&12&40&32&25\\
$NA_{t-1}$ & 0&9&18&0&73&11\\
 \hline
 \end{tabular}\\ 
   \caption{Transition probability matrix (in percentage) of banks between the first quarter of 2011 (indicated with $t-1$) and the second quarter of 2012 (indicated with $t$). The five non-overlapping sets of banks are big borrowers (BB), small borrowers (SB), small lenders (SL), big lenders (BL), and non active (NA). The last column reports the number of banks in each set in the first quarter of 2011.  See text for the definition of the sets. }
\label{transition}
\end{center}
\end{table}

\section{Knock-out numerical experiments}\label{simulations}

We have seen above that some banks have significantly changed their interbank activity around LTRO, either stopping their activity in the market or switching from a borrower to a lender profile (or, much less frequently, viceversa). Here we investigate whether the change of these banks can be responsible of the change in large scale organization of the e-MID market documented in the previous sections.

In order to answer this question we perform knock-out numerical experiments. Similarly to what is done in molecular biology \cite{knock} where genes are silenced or mutated one at a time or in small groups to identify their role in the phenotype, here we drop from the network some banks or modify its in- and out-degree and we perform the inference on the obtained networks to identify banks with a strongest role in preserving the bipartite structure. 

In particular we consider the $12$ monthly-aggregated (binary\footnote{We choose to investigate the binary network because we want to disentangle the change of strategy due to in- and out-degree from the one due to a global variation of the amount of exchanged money.}) networks that display a bipartite structure in 2011 and on the $5$ (binary) monthly-aggregated networks displaying a random structure in 2013, so that we can perform our knock-out experiment on $60$ pairs of banks. We perform two types of modifications to the 2011 network to make it more similar to a 2013 network. The two types of modification are node deletions and degree mutations.  

{\bf Node deletions.} The objective is to test if the disappearance from e-MID of some important bank can be at the origin of the structural change. To this end, for each couple of networks $(A,B)$, where $A$ is a 2011 network and $B$ is a 2013 network, we construct the subnetwork $C$ whose set of nodes is given by the banks active - i.e. with non-zero degree - in both networks, and whose links are given by the corresponding links of the 2011 network $A$. 
For all the 60 couples we find that $C$ maintains the bipartite structure of the starting network $A$, confirming that the change in structure cannot be explained solely by the drop in the number of active banks.

\medskip

{\bf Degree mutations.} The objective here is to test whether the change in strategy (e.g. from borrower to lender) can explain the structural change. To this end  we compute the in-degree $k^i_{in}$ and out-degree $k^i_{out}$ of bank $i$ for each couple of networks $A$ and $B$, and we evaluate for each bank the quantity $\Delta k^i$, which is the Euclidean distance between the 2-component vector $(k^i_{in},k^i_{out})$ in the two networks. This quantity measures the variation in bank's $i$ strategy in the two considered months. Then, starting from the bank with the highest variation $\Delta k$, we substitute its links in $A$ with its links in $B$ and iterate this procedure for all banks in decreasing order of $\Delta k$. Eventually, after enough substitutions, the network will equal $D$ - a subnetwork of $B$ restricted to the nodes in common with $A$. Therefore this substitution procedure starts from $C$ that is bipartite and ends in $D$ that is found to be random. We look for the smallest set of banks with highest $\Delta k$ that must be mutated in order to induce the structural change from bipartite to random. The dimension (number of banks) and composition of this critical set is different in each couple of networks. Therefore we consider the fraction of couples of months in which a bank is in the critical set and thus we obtain a structural score, $S$, for each bank.

\begin{figure}[t]
\centering
\includegraphics[width=80mm]{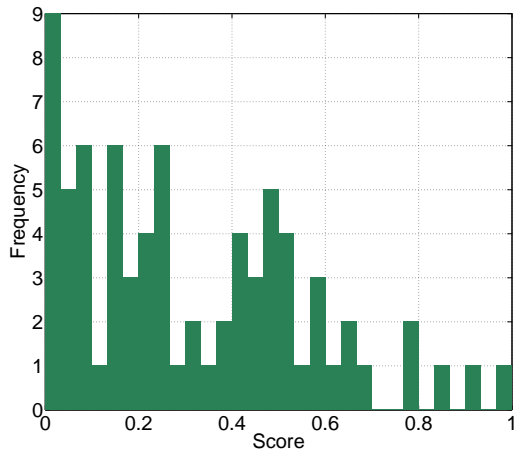}
\caption{Histogram of the score $S$, the fraction of times a bank appears in the critical set for structural change. We see that a small number of banks - right edge - almost always appears in the critical set.}
\label{scoreHisto}
\end{figure}

Figure \ref{scoreHisto} shows the histogram of the scores $S$. It is evident that a small set of banks has a very high score. This observation suggests to consider them {\it structurally important financial institutions}, since the change of their strategy   almost always induces a large scale structural change of the network from bipartite to random. To corroborate this observation we repeat the same substitution procedure, but now we proceed in decreasing order of $S$, i.e. we change the links of the banks with the highest $S$. We find that all the structure changes in the 60 couples of networks can be obtained with the strategy change of the two banks with highest scores, linking the micro-behavior of (structurally) important banks to the macro-organization of the interbank network.

In conclusion the change in e-MID network organization from bipartite to random, observed after LTRO, is likely due to the change of strategy of a very small number of banks. The availability of cheap funding at ECB allowed some banks to change their usual strategy in e-MID, and this change had significant impact on the structure of the interbank network. 

\section{Conclusions}\label{conclusions}

In this paper we propose to use the Stochastic Block Model to investigate and perform a model selection among several possible two block organizations of the network: these include bipartite, core-periphery, and modular structures. Empirically investigating  the e-MID interbank network in the period from January 2010 to December 2014 , we found that a directed bipartite structure best describes the e-MID interbank network at all level of aggregation with the borrower block being more and more densely connected as the aggregation increases. This is different from previous works that assumed the existence of a specific structure (core-periphery) and found the best partition of the set of banks for it. 

The bipartite structure suggests that banks use e-MID market in a strongly polarized fashion depending on their specific needs, borrowing or lending money. Inventory and structural measures also showed the strong persistence in strategies, banks tend to be on the same side of the market (borrower or lender) for several weeks. Only on long time scales (quarterly or longer) the network structure starts to be consistent with core-periphery.  We showed that the ECB extraordinary measures changed the trading behavior of many banks and we argue that, as a consequence of this change of strategy, the structural organization of the network also changed dramatically. While in normal periods (2010-2011 and 2014) bipartite is the most likely structure, in the year after LTRO (2012-2013) a random structure describes better the data. Detailed analysis enabled us to identify also the crucial strategy changes that triggered the structural changes and the role of a very small set of banks in the structural changes. 

All these results contribute to our understanding of the organization of the overnight interbank network as well as of the effect of extraordinary monetary measures on the market. Further investigations are needed from a financial perspective in order to understand how general is the network bipartite structure and the persistence in banks strategies in interbank networks, considering other segments of the interbank network and/or other types of credit (longer maturities and presence of collateral). From a methodological perspective further enquiries might lead to understand the properties and generalize the \textit{knock-out numerical experiment} for the identification of structurally critical banks.

\section*{Acknowledgment}

Authors acknowledge partial support by the grant SNS13LILLB ''Systemic risk in financial markets across time scales".  This work is supported by the European CommunityÕs H2020 Program under the scheme ÔINFRAIA-1-2014-2015: Research InfrastructuresÕ, grant agreement \#654024 ÔSoBigData: Social Mining \& Big Data EcosystemÕ (http://www.sobigdata.eu). Authors also thanks Fabio Caccioli, Thomas Lux, and Daniele Tantari for useful discussions.

\appendix

\section{Algorithms for core-periphery detection}\label{why}

In the literarure the analysis is conducted basically assuming a core-periphery structure for the network and the following different models are used: 
\begin{itemize}
\item The discrete model \cite{FrickeLux}. It involves the minimization of a cost-function associated to each discrete assignment of the nodes. Each link between nodes inside the core lowers the cost-function while each link between nodes inside the periphery raises it, no contribution comes from the links between core nodes and periphery nodes. It has been shown \cite{lip} that this minimization is equivalent to sort nodes in the graph by descending degree, ($k_1>k_2>...>k_N$) and consider the first $s$ nodes, where $s$ is such that $k_s > s+1$. 
Being dependent on the average degree and not on its directionality, this method is insensitive to directed bipartite structures, where there is large directed flow from one group of nodes to another, since nodes with high out-degree and nodes with high in-degree end up inside the core.
\item The symmetric continuous model \cite{FrickeLux}. It involves the minimization of a different cost-function namely:
\begin{equation}
\boldsymbol{u}=\underset{u}{\operatorname{argmin}}\ H_{B}[u]=\underset{u}{\operatorname{argmin}}\sum_{i\neq j}(A_{ij}-u_{i}u_{j})^{2}\label{BOYD}
\end{equation}
Also in this case the symmetry of this coreness measure does not help to understand whether the underlying network has a bipartite structure or a core-periphery structure.
\item The asymmetric continuous model \cite{FrickeLux}. It involves the minimization of a cost-function that distinguishes between an in-coreness and an out-coreness
\begin{equation}
\boldsymbol{u}=\underset{u,v}{\operatorname{argmin}}\ H_{B}[u]=\underset{u}{\operatorname{argmin}}\sum_{i\neq j}(A_{ij}-u_{i}v_{j})^{2}\label{aBOYD}
\end{equation}
This model, as shown in \cite{FrickeLux}, catches the directed structure of the network and indeed significant differences are found between out-coreness and in-coreness. This is a clear indication that a community structure more suitable to grasp the properties of e-MID interbank market is most probably a bipartite one. 
\item The tiering model  \cite{craig}. It can be regarded as an extension of the discrete model, incorporating the role of  asymmetric links and of inter-group links. The basic idea is that a core-node should have both a non-zero in- and out-degree. It relies on a tiering model for banks in a market, the core is a set of intermediary banks that both borrow and lend money to periphery banks.
\end{itemize}

All these methods, especially the directed ones, manage to grasp some of the properties of the e-MID interbank network but a major drawback is that they fix a-priori the block-structure one is supposed to find in the data. In our work we use a well-established model for community detection, the Stochastic Block Model, that allows us to extrapolate the block structure from the data only, without imposing constraints on the mutual relations between groups. As we outline in the text, at all levels of aggregation the unweighted network of e-MID market exhibits a directed bipartite structure, also confirmed by a detailed enquiry of the dynamics of the trading inventory of banks.

\end{document}